%% file: main-arxiv.tex
\documentclass{article}
\usepackage[preprint]{spconf,amsmath,graphicx}
\usepackage{subfig}
\usepackage{booktabs}
\usepackage{amssymb}
\usepackage{caption}
\usepackage{xcolor}
\usepackage{hyperref}

\pdfoutput=1

\copyrightnotice{\begin{tabular}[t]{@{}l@{}}\footnotesize \copyright Copyright 2021 IEEE. Published in ICASSP 2021 - 2021 IEEE International Conference on Acoustics, Speech and Signal Processing (ICASSP), scheduled\\ \footnotesize for 6-11 June 2021 in Toronto, Ontario, Canada.  Personal use of this material is permitted.  However, permission to reprint/republish this material for\\ \footnotesize advertising or promotional purposes or for creating new collective works for resale or redistribution to servers or lists, or to reuse any copyrighted component\\ \footnotesize of this work in other works,  must be obtained from the IEEE. Contact: Manager, Copyrights and Permissions / IEEE Service Center / 445 Hoes Lane / P.O. \\\footnotesize Box 1331 / Piscataway, NJ 08855-1331, USA. Telephone: + Intl. 908-562-3966.\end{tabular}}
\title{A Flow-based neural network for time domain Speech enhancement}
\name{Martin Strauss, Bernd Edler}
\address{International Audio Laboratories Erlangen\textsuperscript{*}\thanks{\textsuperscript{*}A joint institution of the Friedrich-Alexander-University Erlangen-N\"{u}rnberg (FAU) and Fraunhofer IIS, Germany.}, Am Wolfsmantel 33, 91058 Erlangen, Germany\\
\{martin.strauss, bernd.edler\}@audiolabs-erlangen.de}
\begin{document}
\ninept
\maketitle
\begin{abstract}
Speech enhancement involves the distinction of a target speech signal from an intrusive background. Although generative approaches using Variational Autoencoders or Generative Adversarial Networks (GANs) have increasingly been used in recent years, normalizing flow (NF) based systems are still scarse, despite their success in related fields. Thus, in this paper we propose a NF framework to directly model the enhancement process by density estimation of clean speech utterances conditioned on their noisy counterpart. The WaveGlow model from speech synthesis is adapted to enable direct enhancement of noisy utterances in time domain. In addition, we demonstrate that nonlinear input companding benefits the model performance by equalizing the distribution of input samples. Experimental evaluation on a publicly available dataset shows comparable results to current state-of-the-art GAN-based approaches, while surpassing the chosen baselines using objective evaluation metrics.
\end{abstract}
\begin{keywords}
speech enhancement, normalizing flows, deep learning, generative modeling
\end{keywords}
\section{Introduction}
\label{sec:intro}

The goal of speech enhancement (SE) is to emphasize a target speech signal from an interfering background to ensure better intelligibility of the spoken content \cite{Loizou2007}. Due to its importance to a wide range of applications, including, e.g., hearing aids \cite{Borisagar2018} or automatic speech recognition \cite{MOORE2017574}, it has been investigated extensively in the past. In doing so, deep neural networks (DNNs) have largely taken over traditional techniques like Wiener filtering \cite{Lim1978}, spectral subtraction \cite{Boll1979}, subspace methods \cite{Ephraim1995} or the minimum mean square error (MMSE) \cite{Ephraim1985}. Most commonly, DNNs are used to estimate time-frequency (T-F) masks which are able to separate speech and background from a mixture signal \cite{Xu2014}.\\
Nonetheless, systems based on time domain input were proposed in recent years which have the benefit of avoiding expensive T-F transformations \cite{Germain2019, Qian2017, Pascual2017}. Lately, there has also been increasing attention in SE research to generative approaches such as Generative Adversarial Networks (GAN) \cite{Pascual2017, Soni2018, fu2019metricGAN}, Variational Autoencoders (VAE) \cite{Leglaive2019, Pariente2019} and autoregressive models \cite{Qian2017}. Especially the usage of GANs was broadly investigated in the past couple of years. For instance, Pascual et al. \cite{Pascual2017}, proposed a GAN-based end-to-end system, where the generator enhances noisy speech samples directly on a waveform level. This approach has been extended multiple times, e.g. by making use of the Wasserstein distance \cite{Adiga2019} or by combining multiple generators to increase performance \cite{Phan2020}. Others reported strong SE results working with GANs to estimate clean T-F spectrograms by implementing additional techniques like a mean squared error regularization \cite{Soni2018} or optimizing the network directly with respect to a speech specific evaluation metric \cite{fu2019metricGAN}.\\
While the aforementioned approaches received increasing popularity lately, normalizing flow (NF) based systems are still rare in SE. Just recently, the work of Nugraha et al. \cite{Nugraha2020} proposed a flow based model combined with a VAE to learn a deep latent representation, which can be used as a prior distribution for speech. Yet, their approach does not model the enhancement process itself and therefore depends on the SE algorithm it is combined with. However, it was shown in areas like computer vision \cite{ho19a} or speech synthesis \cite{Prenger2019} that NFs have the ability to successfully generate high quality samples in their respective tasks. Consequently, this leads to the assumption that enhancement of speech samples can be performed directly using a flow-based system by modeling a generative process.\\
In this paper we demonstrate that NFs can successfully be applied to SE by a learned mapping from a probability distribution based on clean speech samples conditioned on their noisy counterpart to a Gaussian distribution. Therefore, we modify a state-of-the-art flow-based DNN architecture from speech synthesis to perform SE directly in time domain, without the need of any predefined features or T-F transformations. Further, we show that an easy preprocessing technique of the input signal using companding can increase the performance of SE models based on density estimation. The experimental evaluation of the proposed methods confirms our assumptions and shows comparable results to current state-of-the-art systems, while outperforming other time domain GAN baselines.

\begin{figure}[htb]
\begin{minipage}[b]{1.0\linewidth}
  \centering
  \def\svgwidth{8.5cm}
  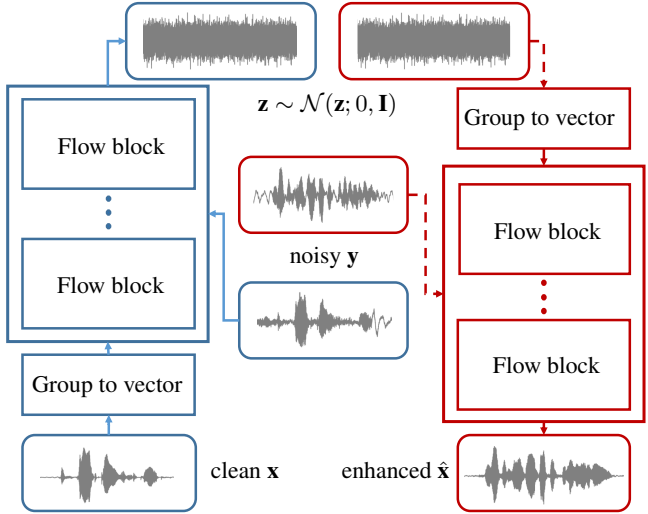
\end{minipage}
\caption{Overview of the proposed system. During training the clean-noisy ($\textbf{x}$-$\textbf{y}$) pair follows the flow blocks (blue-solid lines) to be mapped to a Gaussian distribution, $\mathcal{N}(\textbf{z}; 0, \textbf{I})$. In inference, a sample $\textbf{z}$ is drawn from this distribution and follows the inverted flow (red-dashed lines) together with another noisy utterance $\textbf{y}$ to generate the enhanced signal $\hat{\textbf{x}}$ (Best viewed in colors).}
\label{fig:arch}
\end{figure}

\section{Problem Formulation}
\label{sec:problem_description}
\subsection{Normalizing flow fundamentals}
A normalizing flow is a differentiable invertible transformation with differentiable inverse \cite{Kobyzev2020}, formally expressed as

\begin{equation}
    \textbf{x} = f(\textbf{z}),
\end{equation}

where $\textbf{x} \in \mathbb{R}^{D}$ and $\textbf{z} \in \mathbb{R}^{D}$ are $D$ dimensional random variables and $f$ is the function mapping $\textbf{z}$ to $\textbf{x}$. Given that $f$ is invertible, i.e.,

\begin{equation}
    \textbf{z} = f^{-1}(\textbf{x}),
\end{equation}

it is ensured, that $\textbf{x}$ is a random variable with a given probability distribution. Following this, the log probability density function $\log  p_x(\textbf{x})$ can be computed by a change of variables \cite{Papamakarios2019NormalizingFF},

\begin{equation}
    \log p_x(\textbf{x}) = \log p_z(f^{-1}(\textbf{x})) + \log \left\lvert \det \left( J(\textbf{x}) \right) \right\rvert ,
\end{equation}

where ${J(\textbf{x})=\partial f^{-1}(\textbf{x}) / \partial \textbf{x}}$ defines the Jacobian consisting of all first order partial derivatives. Moreover, if every single transformation in a  sequence of $1$ to $T$ transformations is invertible and differentiable, then the same is true for their composition, i.e.,

\begin{equation}
    \textbf{x} = f_1 \circ f_2 \circ \cdots \circ f_T(\textbf{z}).
\end{equation}

\subsection{Speech enhancement flow}
In the case of speech enhancement we consider a time domain mixture signal $\textbf{y} \in \mathbb{R}^{1 \times N}$ of length $N$ composed of a clean speech utterance $\textbf{x} \in \mathbb{R}^{1 \times N}$ and some additive disturbing background $\textbf{n} \in \mathbb{R}^{1 \times N}$ so that

\begin{equation}
    \textbf{y}=\textbf{x}+\textbf{n}.
\end{equation}

Moreover, we define

\begin{equation}
    \textbf{z} \sim \mathcal{N}(\textbf{z}; 0,\textbf{I}),
\end{equation}

where $\mathcal{N}$ is a Gaussian distribution with zero mean and unit variance. The NF is now modelled by a DNN and aims to outline the probability distribution $p_x(\textbf{x}|\textbf{y})$ of clean speech utterances $\textbf{x}$ conditioned on the noisy mixture $\textbf{y}$. As training objective the log-likelihood of the previously defined probability distribution is maximized, i.e.,

\begin{equation}
    \log p_x(\textbf{x}|\textbf{y}; \theta) = \log p_z(f^{-1}_{\theta}(\textbf{x})|\textbf{y}) + \log \left\lvert \det \left( J(\textbf{x}) \right) \right\rvert,
\end{equation}

where $\theta$ represents the network parameters. See Figure \ref{fig:arch} for an illustration of the training process (left, blue lines). In the enhancement step, we can sample from $p_z(\textbf{z})$ and hand it together with a noisy speech sample as input to the network. Following the inverted flow the DNN maps the random sample back to the distribution of clean utterances to create $\hat{\textbf{x}}$, with $\hat{\textbf{x}}$ ideally being close to the underlying $\textbf{x}$. This process is also displayed in Figure \ref{fig:arch} (right, red lines).

\section{Proposed methods}
\label{sec:methods}
\subsection{Model architecture}
\label{ssec:model_arc}
In this work, we adapted the WaveGlow architecture \cite{Prenger2019} for speech synthesis to speech enhancement. Originally, this model takes speech utterances together with a corresponding Mel-spectrogram as an input for several flow blocks, learning to generate realistic speech samples based on the conditional spectrogram input. One flow block consists of a 1x1 invertible convolution \cite{glow2018} ensuring the exchange of information along channel dimension and a so-called affine coupling layer \cite{DinhSB17}, which is used to ensure invertibility and efficient computing of the Jacobian determinant. Therefore, the input signal is split into two halves along the channel dimension. One half is then fed to a WaveNet-like DNN block defining the scaling and translation factor for the second half. To create this multi-channel input, multiple audio samples are stacked together in one group to mimic a multi-channel signal. The affine coupling layer is also where the conditional information is included. For further details on this procedure we refer to \cite{Prenger2019}.\\
The original WaveGlow is computationally heavy ($>87$\,Mio. parameters), so a few architectural modifications were carried out to make it feasible to train on a single GPU and to enable the enhancement of speech samples. In contrast to WaveGlow, we condition the model on noisy time domain speech signals, instead of Mel-spectrograms. Since both signals are of the same dimension, no upsampling layer was needed. Additionally, the standard convolutions in the WaveNet-like blocks were replaced by depthwise separable convolutions \cite{chollet2017} to reduce the amount of parameters, as it was proposed by \cite{squeezewave}.

\subsection{Nonlinear input companding}

\setlength{\tabcolsep}{15pt}
\begin{table*}[ht]
\renewcommand{\arraystretch}{1.2}
\centering
\caption{Evaluation results using objective evaluation metrics. SE-Flow represents the proposed flow-based approach and SE-Flow-$\mu$ the approach together with $\mu$-law companding of the input data. The values of all comparing methods are taken from the corresponding papers.}
\begin{tabular}{llllll}\toprule
Method & \multicolumn{1}{r}{PESQ} & \multicolumn{1}{r}{CSIG} & \multicolumn{1}{r}{CBAK} & \multicolumn{1}{r}{COVL} & \multicolumn{1}{r}{segSNR} \\ \midrule
Noisy  & \multicolumn{1}{r}{1.97} & \multicolumn{1}{r}{3.35} & \multicolumn{1}{r}{2.44} & \multicolumn{1}{r}{2.63} & \multicolumn{1}{r}{1.68} \\ \midrule
SEGAN \cite{Pascual2017} & \multicolumn{1}{r}{2.16} & \multicolumn{1}{r}{3.48} &\multicolumn{1}{r}{2.94} & \multicolumn{1}{r}{2.80} & \multicolumn{1}{r}{7.73} \\ \midrule
DSEGAN \cite{Phan2020} & \multicolumn{1}{r}{2.39} & \multicolumn{1}{r}{3.46} &\multicolumn{1}{r}{3.11} & \multicolumn{1}{r}{3.50} & \multicolumn{1}{r}{-} \\ \midrule
MMSE-GAN \cite{Soni2018}  &\multicolumn{1}{r}{2.53}  & \multicolumn{1}{r}{3.80} & \multicolumn{1}{r}{3.12} & \multicolumn{1}{r}{3.14} & \multicolumn{1}{r}{-} \\\midrule
Metric-GAN \cite{fu2019metricGAN}  &\multicolumn{1}{r}{2.86}  & \multicolumn{1}{r}{3.99} & \multicolumn{1}{r}{3.18} & \multicolumn{1}{r}{3.42} & \multicolumn{1}{r}{-} \\\midrule \midrule
SE-Flow    &\multicolumn{1}{r}{2.28}  & \multicolumn{1}{r}{3.70} & \multicolumn{1}{r}{3.03} & \multicolumn{1}{r}{2.97} & \multicolumn{1}{r}{7.93} \\\midrule
SE-Flow-$\mu$ & \multicolumn{1}{r}{2.43}  & \multicolumn{1}{r}{3.77} &\multicolumn{1}{r}{3.12}  & \multicolumn{1}{r}{3.09} &  \multicolumn{1}{r}{8.07} \\
       \bottomrule
\end{tabular}
\vspace{-0.2cm}
\label{tab:results}
\end{table*}

\begin{figure}[htb]
\begin{minipage}[b]{1.0\linewidth}
  \centering
  \centerline{\includegraphics[width=8.5cm]{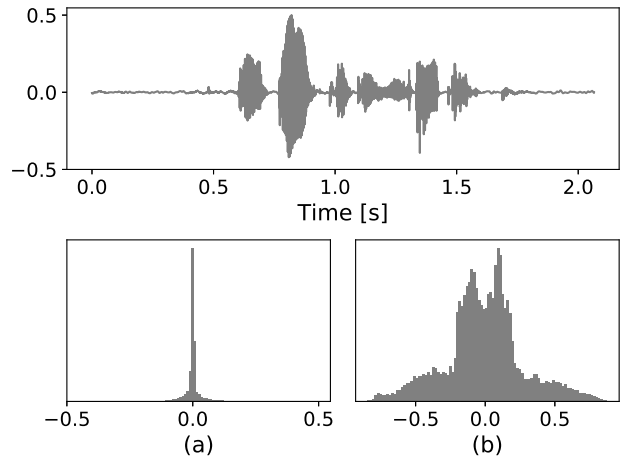}}
\end{minipage}
\caption{Example of the effect of nonlinear $\mu$-law companding on the audio input. At the top a clean speech utterance is shown. (a) shows the histogram ($n_{bins}=100$) of the clean utterance. The companded values are illustrated in (b).}

\label{fig:companding}
\end{figure}

Since the DNN models the probability distribution of time domain speech utterances, it is important to inspect the range of values from which the distribution is learned from. The audio data was stored as normalized 32-bit floats in a range of $[-1,1]$. Since time domain speech samples approximately follow a Laplacian distribution \cite{jensen2005study}, it is easy to see that most values lie in a small range around zero (See Figure \ref{fig:companding} (a)). However, especially in clean speech utterances, data samples with a higher absolute amplitude carry significant information and are in this case underrepresented. To make sure that the values are more evenly spread out, a nonlinear companding algorithm can be applied to map small amplitudes to a wider interval and large amplitudes to a smaller interval, respectively. This is demonstrated in Figure \ref{fig:companding}, where one speech sample is displayed together with a histogram of values before and after applying a companding algorithm. In this sense, the companding can be understood as a sort of histogram equalization. Following this, additional experiments  were conducted using $\mu$-law companding (ITU-T. Recommendation G. 711) of the input data without quantization, which is formally defined as

\begin{equation}
    g(x) = \text{sgn}(x)\frac{\ln (1+\mu\left\lvert x \right\rvert)}{\ln (1+\mu)},
\end{equation}

where $\text{sgn()}$ is the sign function and $\mu$ is the parameter defining the level of compression. Here, we set $\mu$ to 255 throughout the experiments, which is a common value also used in telecommunication. With respect to the learning objective, rather than to learn the distribution of clean utterances, the distribution of the companded signal is learned. The enhanced sample can be expanded back to a regular signal afterwards reverting the $\mu$-law transformation.

\section{Experimental Setup}
\label{sec:experiments}

\subsection{Data}
\label{ssec:subhead}

\begin{figure*}[htb]
  \centering
  \subfloat[\centering Noisy]{\includegraphics[width=4.52cm]{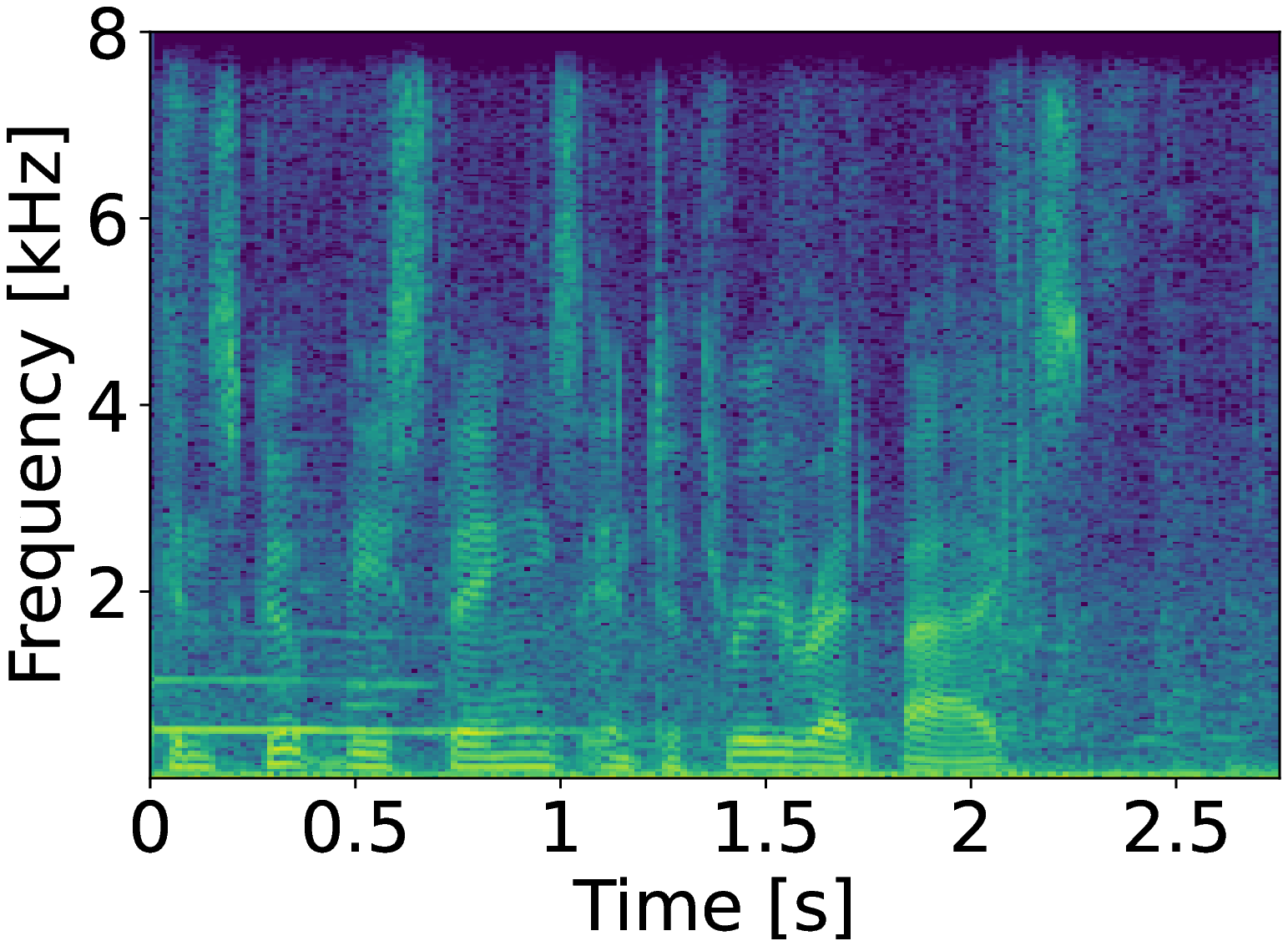}}
    \subfloat[\centering Clean]{\includegraphics[width=4.42cm]{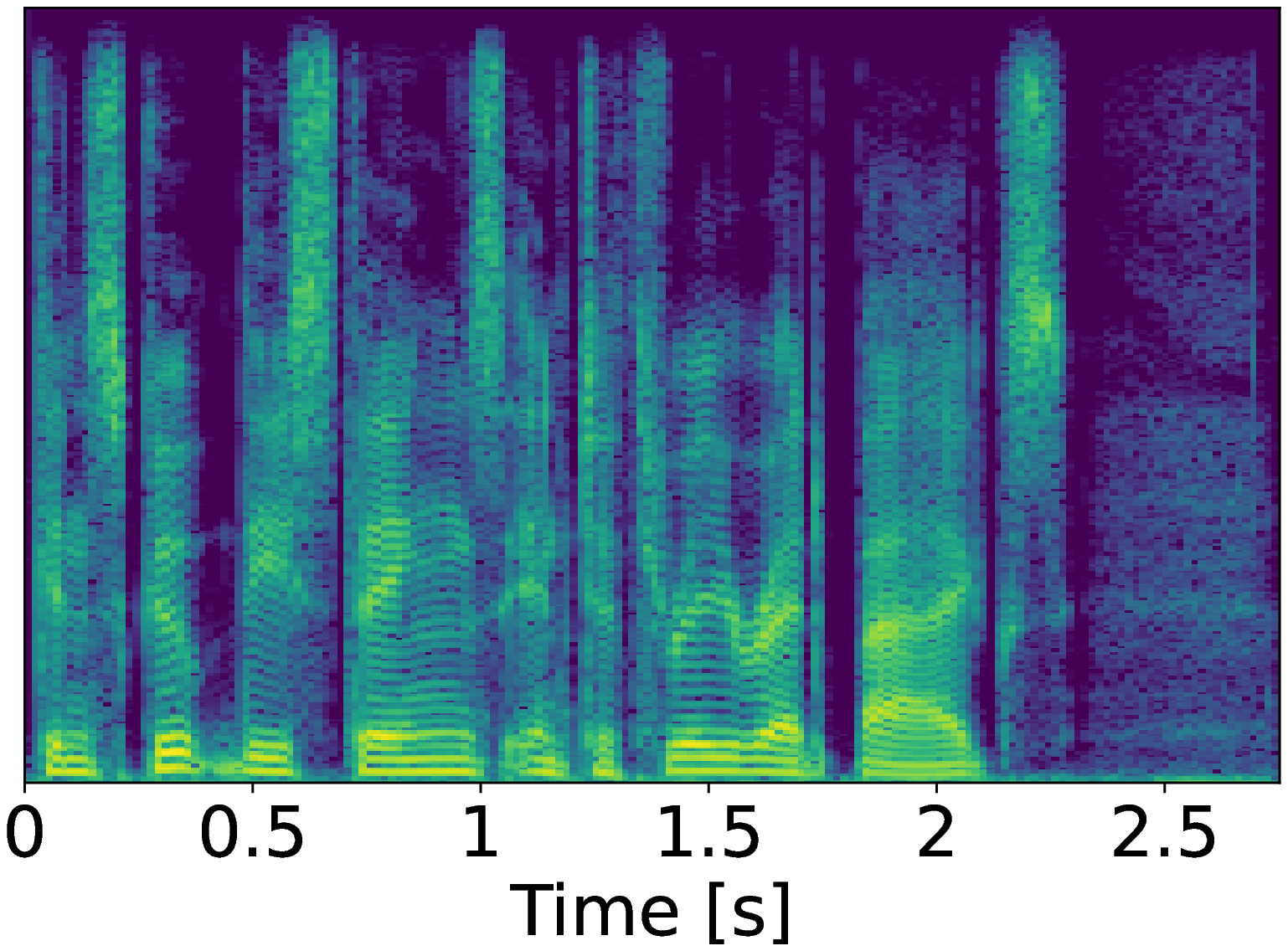}}
    \subfloat[\centering SE-Flow]{\includegraphics[width=4.42cm]{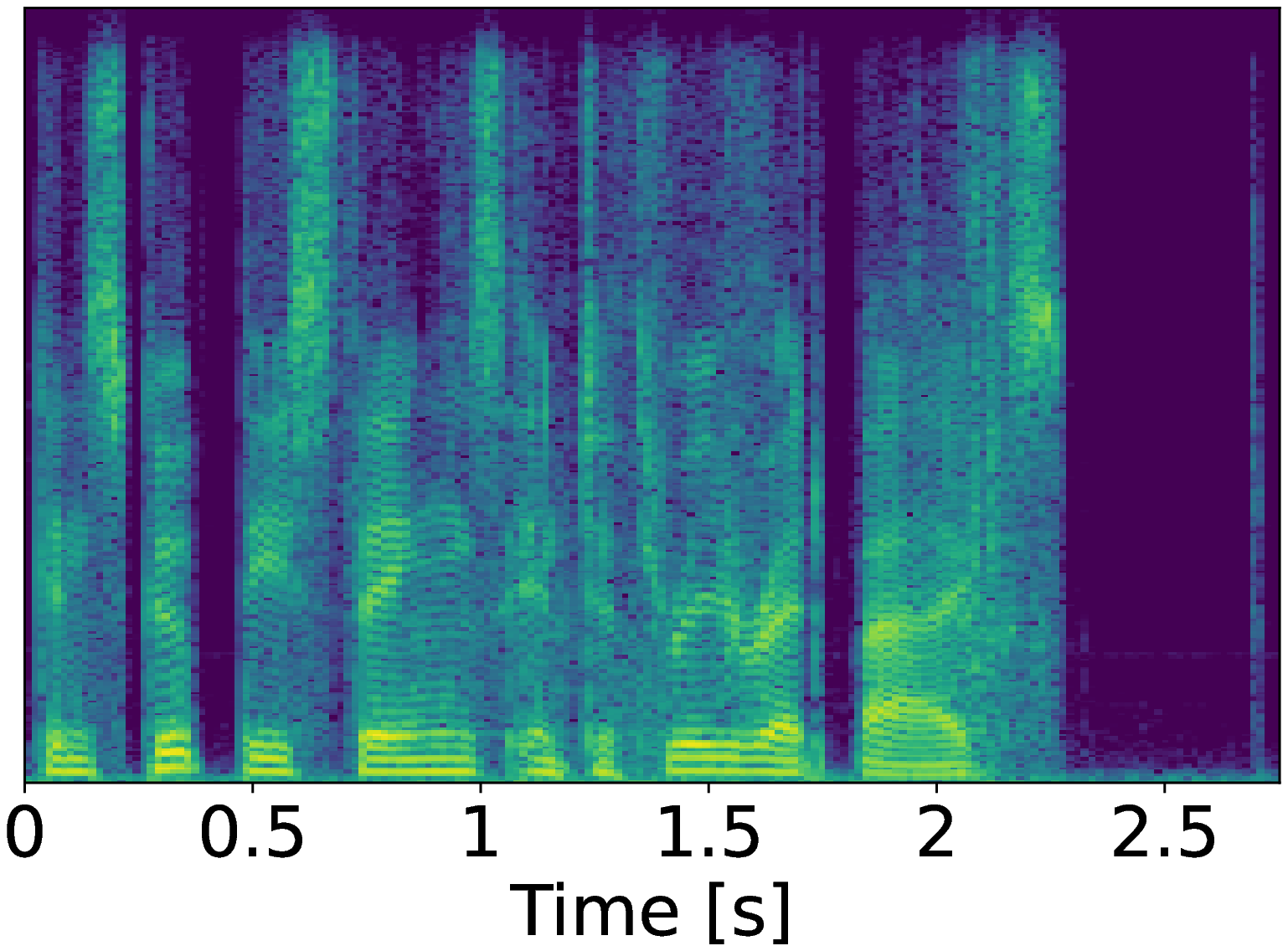}}
    \subfloat[\centering SE-Flow-$\mu$]{\includegraphics[width=4.42cm]{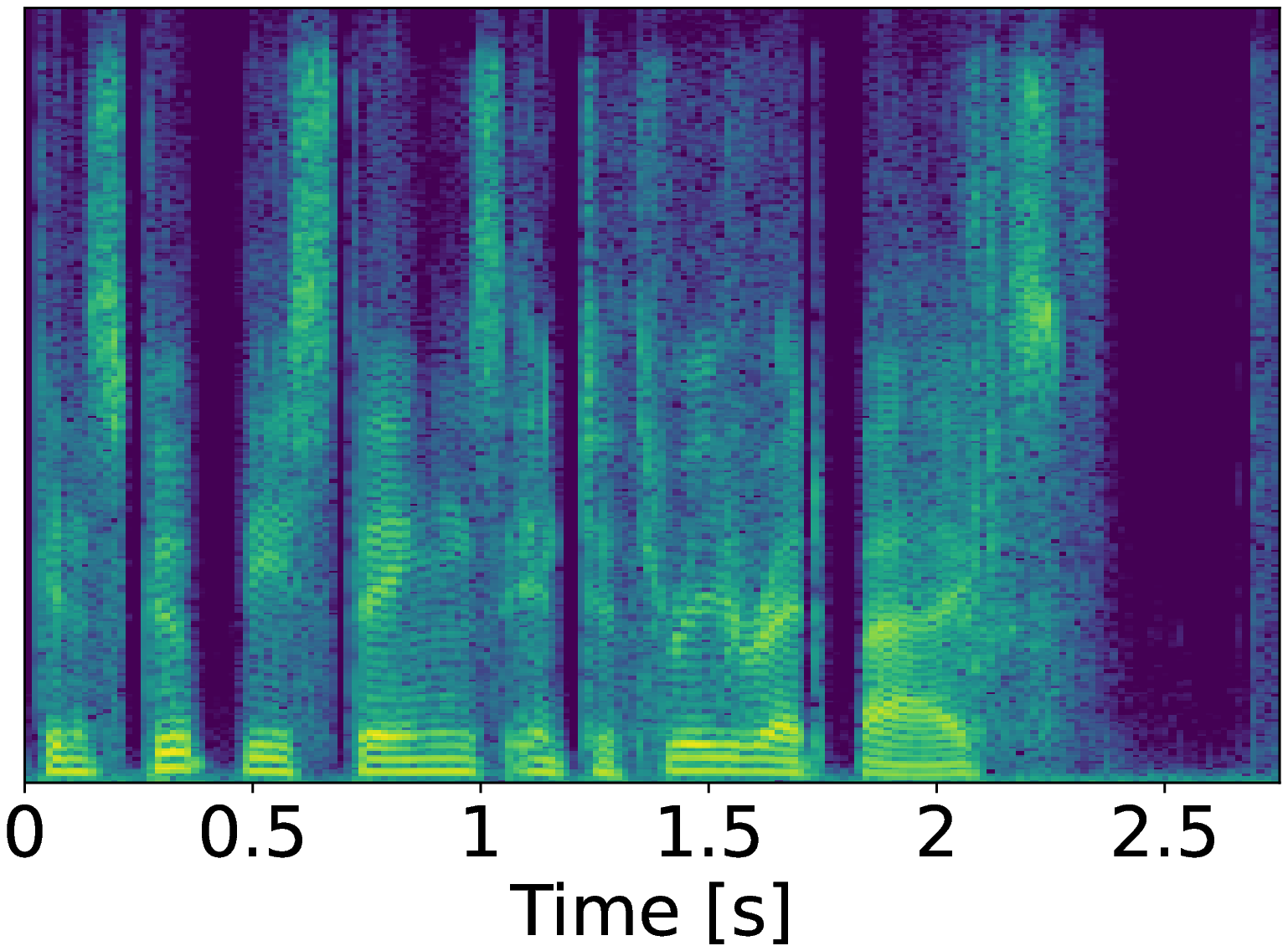}}
  \caption{Spectrograms obtained from proposed the system. In (a), a noisy speech utterance at 2.5\,dB is displayed. (b) shows the corresponding clean utterance. In (c) and (d) the results of the proposed flow-based systems are shown.}
  \vspace{-0.3cm}
  \label{fig:spec}
\end{figure*}

The dataset used in this study was published along with the work of Valentini et al. \cite{valentini2016}. It includes 30 individual speakers from the Voice Bank corpus \cite{voicebank} separated into a training and a test set with 28 and 2 speakers, respectively. Both sets are balanced according to male and female participants. The training samples were mixed together with eight real noise samples from the DEMAND database \cite{demand} and two artificial (babble and speech shaped) samples according to 0, 5, 10 and 15\,dB signal-to-noise-ratio (SNR). In addition, one male and one female speaker were taken out of the training set to form a validation set for model development, resulting in 8.6\,h and 0.7\,h of data for training and validation, respectively. To ensure unseen test conditions, different noise samples were selected and mixed according to SNR values of 2.5, 7.5, 12.5, and 17.5\,dB to form 824 (0.6\,h) test utterances. All audio files were resampled to a sampling rate of $f_s=16$\,kHz.

\subsection{Training strategy}
The values for batch size ($4, 8, 12$), number of flow blocks ($8, 12, 16$) and the amount of samples grouped together as input ($8, 12, 24$) were selected in a hyper-parameter search. As a loss function for training, the model minimized the negative log-likelihood of the given data. Each individual model was trained for 150 epochs to select the parameters based on the lowest validation loss. A few initial experiments using a higher batch size were conducted, however we found that the model was not generalizing well enough. The selected model was trained further until convergence based on an early stopping mechanism of 20 epochs patience. A fine-tuning step followed using a decreased learning rate and the same early stopping criterion.

\subsection{Model settings}
As a result of the parameter search the model was built with 16 flow blocks, a group of 12 samples as input and a batch size of 4. The learning rate was set to $3\times10^{-4}$ in the initial training step using the Adam \cite{KingmaB14} optimizer and weight normalization \cite{NIPS2016_6114}. For fine-tuning the learning rate was lowered to  $3\times10^{-5}$.  The input at training time consisted of 1\,s long randomly extracted chunks of each audio file. The standard deviation of the Gaussian distribution was set to $\sigma=1.0$. Similar to other NF models \cite{Prenger2019}, we observed the effect that using a lower value for $\sigma$ in inference leads to a  higher quality output, which is why it was set to $\sigma=0.9$ for the enhancement. According to the original WaveGlow architecture WaveNet-like blocks with 8 layers of dilated convolutions, 512 channels as residual connections and 256 channels in the skip connections were used in the affine coupling layers. In addition, after every 4 coupling layers 2 channels were passed to the loss function to form a multi-scale architecture.

\subsection{Evaluation}
In order to compare the proposed approach to recent works in the field the following evaluation metrics were used:
\begin{itemize}
    \item Perceptual  evaluation  of  speech  quality (PESQ) in the recommended wideband version in ITU-T P.862.2 (from $-0.5$ to $4.5$).
    \item Three Mean Opinion Score (from $1$ to $5$)  metrics \cite{Hu2008}: prediction of the signal distortion (CSIG), prediction of the background intrusiveness (CBAK) and prediction of the overall speech quality (COVL).
    \item The segmental SNR (segSNR) \cite{Hansen1998AnEQ} improvement (from $0$ to $\infty$\,dB).
\end{itemize}

As a baseline to our proposed methods we used two generative time domain approaches, namely the SEGAN \cite{Pascual2017} and the improved deep SEGAN (DSEGAN) \cite{Phan2020} models, since they were evaluated with the same database and metrics. Further, we compared against two other state-of-the-art GAN based systems, namely the MMSE-GAN \cite{Soni2018} and Metric-GAN \cite{fu2019metricGAN}, which are working with T-F masks. Note, that there are several discriminative  approaches, e.g.  \cite{Germain2019, Giri2019, Koizumi2020}, reporting a higher performance on this dataset. However, the focus of this work was on generative models, which is why they are not included in the comparison.

\section{Experimental Results}
\label{results}
The experimental results are displayed in Table \ref{tab:results}. As is shown in the table, in between the two proposed flow-based experiments, the model using $\mu$-companding shows better results in all metrics. This demonstrates the effectiveness of this easy preprocessing technique for modeling the distribution of time domain signals. An illustration of the enhancement capabilities is illustrated in Figure \ref{fig:spec}. Comparing the spectrograms of the two proposed systems, it appears that the SE-Flow-$\mu$ is able to capture more fine grained speech parts with less background leakage. Note also, that the breathing sound at the end of the displayed example is not recovered by our models, which emphasizes that the proposed models focuses on real speech samples. Further, it is visible in the flow-based examples that, when speech is active, more noise-like frequency content is present in the higher frequencies compared to the clean signal. This can be explained by the Gaussian sampling which is not modelled properly during inference.\\
In comparison to the SEGAN baseline both proposed methods show superior performance throughout all metrics by a large margin.  Looking at DSEGAN it can be seen that the proposed SE-Flow reaches a comparable performance in CSIG, while showing slightly lower values in the other metrics. However, the SE-Flow-$\mu$ based system still performs better in all metrics besides COVL. Thus, within the time domain approaches the proposed flow-based model seems to model the generative process from a noisy to an enhanced signal better. With regard to MMSE-GAN, we observe that our approach with $\mu$-law companding has a similar performance with a slight edge towards MMSE-GAN, although no additional regularization technique was implemented here. The Metric-GAN shows superior results compared to the proposed approaches with regard to all displayed metrics. Still, it is important to notice that this model was directly optimized according to the PESQ metric, so we would expect a good performance here. Consequently, connecting the training with direct optimization of the evaluation metric might also be an effective way to improve our system. Samples of the enhanced signals produced by our models can be found online\footnote[1]{\url{https://www.audiolabs-erlangen.de/resources/2021-ICASSP-SE\_Flow}}.

\section{Conclusions}
\label{sec:conclusion}
In this paper we proposed a normalizing flow method for speech enhancement. The model allows for density estimation of clean speech samples given their noisy counterparts and signal enhancement via generative inference. A simple nonlinear $\mu$-law companding technique was demonstrated to be an effective preprocessing tool to increase the enhancement outcome. The proposed systems outperforms the results of other time domain GAN-based baselines while closing up to state-of-the-art T-F techniques. Future research will include the exploration of different techniques in the coupling layer, as well as, a combination of time and frequency domain signals.

\section{Acknowledgments}
We want to thank Dr.\,Nicola Pia and Wolfgang Mack for the thoughtful comments and constructive feedback in the early and late stages of our work.



\end{document}

%% file: architecture.pdf_tex
\begingroup%
  \makeatletter%
  \providecommand\color[2][]{%
    \errmessage{(Inkscape) Color is used for the text in Inkscape, but the package 'color.sty' is not loaded}%
    \renewcommand\color[2][]{}%
  }%
  \providecommand\transparent[1]{%
    \errmessage{(Inkscape) Transparency is used (non-zero) for the text in Inkscape, but the package 'transparent.sty' is not loaded}%
    \renewcommand\transparent[1]{}%
  }%
  \providecommand\rotatebox[2]{#2}%
  \newcommand*\fsize{\dimexpr\f@size pt\relax}%
  \newcommand*\lineheight[1]{\fontsize{\fsize}{#1\fsize}\selectfont}%
  \ifx\svgwidth\undefined%
    \setlength{\unitlength}{669.21103585bp}%
    \ifx\svgscale\undefined%
      \relax%
    \else%
      \setlength{\unitlength}{\unitlength * \real{\svgscale}}%
    \fi%
  \else%
    \setlength{\unitlength}{\svgwidth}%
  \fi%
  \global\let\svgwidth\undefined%
  \global\let\svgscale\undefined%
  \makeatother%
  \begin{picture}(1,0.79876685)%
    \lineheight{1}%
    \setlength\tabcolsep{0pt}%
    \put(0,0){\includegraphics[width=\unitlength,page=1]{architecture.pdf}}%
    \put(0.08082403,0.56111901){\color[rgb]{0,0,0}\makebox(0,0)[lt]{\lineheight{1.25}\smash{\begin{tabular}[t]{l}Flow block\end{tabular}}}}%
    \put(0.07972383,0.34290996){\color[rgb]{0,0,0}\makebox(0,0)[lt]{\lineheight{1.25}\smash{\begin{tabular}[t]{l}Flow block\end{tabular}}}}%
    \put(0.76302833,0.4274191){\color[rgb]{0,0,0}\makebox(0,0)[lt]{\lineheight{1.25}\smash{\begin{tabular}[t]{l}Flow block\end{tabular}}}}%
    \put(0.76131132,0.21742362){\color[rgb]{0,0,0}\makebox(0,0)[lt]{\lineheight{1.25}\smash{\begin{tabular}[t]{l}Flow block\end{tabular}}}}%
    \put(0.71652213,0.60569649){\color[rgb]{0,0,0}\makebox(0,0)[lt]{\lineheight{1.25}\smash{\begin{tabular}[t]{l}Group to vector\end{tabular}}}}%
    \put(0.04061341,0.18803929){\color[rgb]{0,0,0}\makebox(0,0)[lt]{\lineheight{1.25}\smash{\begin{tabular}[t]{l}Group to vector\end{tabular}}}}%
    \put(0.32024083,0.05340245){\color[rgb]{0,0,0}\makebox(0,0)[lt]{\lineheight{1.25}\smash{\begin{tabular}[t]{l}clean $\textbf{x}$\end{tabular}}}}%
    \put(0.44344566,0.39014223){\color[rgb]{0,0,0}\makebox(0,0)[lt]{\lineheight{1.25}\smash{\begin{tabular}[t]{l}noisy $\textbf{y}$\end{tabular}}}}%
    \put(0.52489644,0.0529899){\color[rgb]{0,0,0}\makebox(0,0)[lt]{\lineheight{1.25}\smash{\begin{tabular}[t]{l}enhanced $\hat{\textbf{x}}$\end{tabular}}}}%
    \put(0.39347443,0.62850977){\color[rgb]{0,0,0}\makebox(0,0)[lt]{\lineheight{1.25}\smash{\begin{tabular}[t]{l}$\textbf{z} \sim \mathcal{N}(\textbf{z}; 0,\textbf{I})$\end{tabular}}}}%
  \end{picture}%
\endgroup%